%% file: main.tex
\documentclass[10pt,letterpaper,compsoc,conference]{iiswc26}

\usepackage{cite}
\usepackage{amsmath,amsfonts}
\usepackage{algorithmic}
\usepackage{graphicx}
\usepackage[dvipsnames]{xcolor}
\usepackage[final]{microtype}
\usepackage[italic]{mathastext}
\usepackage{libertine}
\usepackage[T1]{fontenc}
\usepackage{textcomp}
\usepackage[varqu,varl]{zi4}
\usepackage[all]{nowidow}
\usepackage[auth-lg,affil-it]{authblk}
\usepackage[keeplastbox]{flushend}
\usepackage{fancyhdr}

\usepackage{enumitem}
\usepackage{hyperref}

\usepackage{orcidlink}

\newcommand{\costDeployPrimeFactorizationfsze}{237,898,716}
\newcommand{\costDeployOrderFindingfsze}{6,370,521,379}

\newcommand{\costSolveOrderFindingMinfsze}{141,650}
\newcommand{\costSolveOrderFindingMaxfsze}{16,074,742}
\newcommand{\costSolveOrderFindingMeanfsze}{8,108,235}
\newcommand{\costSolveOrderFindingMedianfsze}{8,122,201}

\newcommand{\costSolvePrimeFactorizationAllfsze}{741,048,802}
\newcommand{\costSolvePrimeFactorizationOnefsze}{6,110,290}
\newcommand{\bountyGasfsze}{800,000,000}

\newcommand{\modulusBitSizefsze}{4608}
\newcommand{\modulusBitSizeMinusOnefsze}{4607}
\newcommand{\modulusBitSizeWithCommafsze}{4,608}
\newcommand{\maxOrderfsze}{1,152}
\newcommand{\bitSizeOfPrimesfsze}{1536}

\fancypagestyle{firstpage}{
  \fancyhf{}
  
  \fancyhead[C]{} 
  \fancyfoot[C]{\thepage}
}

\begin{document}

%% EDIT TITLE BELOW

\title{BloQBench: A Blockchain Benchmarking Framework for Quantum Supremacy}

%% DO NOT EDIT THE FOLLOWING

%\renewcommand\Authsep{\qquad}
%\renewcommand\Authand{\qquad}
%\renewcommand\Authands{\qquad}

%% EDIT AUTHOR LIST BELOW

\author{Nicholas J.C. Papadopoulos\orcidlink{0000-0002-6357-0030}}
\author{Ramin Ayanzadeh\orcidlink{0000-0001-6687-5668}}
\affil{University of Colorado Boulder}

%%% ALTERNATIVE FORMAT FOR MULTIPLE SCHOOLS:
%%% 
% \author[1]{Author1 Name}
% \author[2]{Author2 Name}
% \author[2]{Author3 Name}
% \author[1]{Author4 Name}
% \affil[1]{Full Name of Awesome School}
% \affil[2]{Full Name of Awesomer School}

\maketitle
\thispagestyle{firstpage}
\pagestyle{plain}

%% EDIT YOUR PAPER'S CONTENTS BELOW

\input{./sections/0abstract}
\input{./sections/1introduction}
\input{./sections/background_and_motivation}
\input{./sections/2result}
\input{./sections/3discussion}
\input{./sections/4methods}
\input{./sections/5conclusion}

\section*{Acknowledgments}
During the preparation of this work, the author(s) used generative AI/LLM tools to improve the grammar and readability of the manuscript. After using this tool, the author(s) reviewed and edited the content as needed and take full responsibility for the final publication.

We thank Danny Ryan (GitHub @djrtwo) for his valuable contributions to the development of this contract, particularly through insightful discussions and conceptual input throughout the process.
Flowcharts were created in Lucid (lucid.co).

\bibliographystyle{IEEEtranS}
\bibliography{references}

\clearpage
\input{./sections/artifact}

\end{document}

%% file: sections/0abstract.tex
\begin{abstract}
As quantum computing matures, characterizing its practical workloads and verifying quantum supremacy presents a significant challenge. 
Current benchmarking and claims rely on trust-based verification methods that lack public auditability.
We propose a decentralized benchmarking framework implemented via an Ethereum smart contract, to provide verifiable assurance in these claims.
This framework generates classically intractable puzzles that, crucially, require absolutely no pre-computed secrets.
By utilizing the blockchain as an immutable public ledger, independent observers can mathematically verify that any provided solution to the puzzle must have been computationally derived via quantum hardware rather than classically spoofed.
Furthermore, we demonstrate how this verifiable benchmarking metric can be utilized as an automation trigger.
As a practical example of such a trigger, we focus on the ability for blockchains to automatically switch to quantum-secure signature schemes upon the successful demonstration of cryptographic quantum supremacy. 
We demonstrate these principles with \emph{BloQBench}, which implements the concept using integer factorization as the generated puzzle and Lamport signatures as the trigger-based effect. 
This approach demonstrates a novel use of distributed ledgers for quantum workload characterization, providing a transparent, automated metric for measuring quantum supremacy while managing the performance and complexity trade-offs of post-quantum technology transitions.
\end{abstract}

%% file: sections/1introduction.tex
\section{Introduction}

As quantum computing capabilities advance, the ability to accurately characterize their practical workloads and verify claims of quantum advantage has become a critical challenge~\cite{preskill2012quantumcomputingentanglementfrontier}.
Recent efforts to demonstrate quantum supremacy, typically through highly contrived experiments ~\cite{kim,arute,morvan2023phasetransitionrandomcircuit,Titcomb_2023,Neven_2024}, have frequently been met with rebuttals demonstrating that classical supercomputers can perform these computations far more efficiently than originally anticipated, undermining the claimed quantum advantage~\cite{begušić2023fastclassicalsimulationevidence,Pednault2019LeveragingSS,Pednault_Maslov_Gunnels_Gambetta_2019b}. 
This back-and-forth highlights a fundamental bottleneck in quantum workload characterization, as there is no trustless and agreed-upon means of proving or verifying quantum supremacy.
For a workload to definitively demonstrate quantum supremacy, it must be computationally intractable for classical hardware to solve. 
However, if verifying the correct execution of that workload is also classically intractable, independent validation becomes practically impossible. 
The community is, therefore, increasingly forced to rely on localized, trust-based verification methods, lacking a universally auditable standard to prove that a specific workload was executed on quantum hardware and is intractable for classical hardware.

We address this verification issue with workloads that are inherently difficult to compute but trivial to verify.
While integer factorization via Shor's algorithm~\cite{shor_97,improvedFactoring} provides such an asymmetry, relying on trusted third parties to generate these workloads introduces potential bias, as the generator may harbor pre-computed secrets.
Generating a product of large primes, for example, generally requires first generating and multiplying secret primes, then discarding them.
If these foundational primes are retained or leaked, an actor could trivially submit the solution to spoof the benchmark without ever executing the quantum workload. 
Consequently, this approach necessitates an underlying trust in the workload generator, which is unacceptable when evaluating definitive claims of quantum supremacy.

We establish a novel paradigm for quantum performance evaluation by introducing the first entirely trustless, publicly auditable benchmarking framework deployed directly on a distributed ledger.
While blockchains like Bitcoin~\cite{bitcoin_whitepaper} and Ethereum~\cite{ethereum} are traditionally viewed through the lens of decentralized finance and cryptocurrency~\cite{cryptocurrencies}, they effectively function as globally distributed, immutable execution environments. 
Ethereum, in particular, supports smart contracts~\cite{smart_contracts}, which are deterministic programs executed across a decentralized network where all state transitions are publicly verifiable. 
By utilizing a smart contract to probabilistically generate classically intractable factorization puzzles without any secret parameters, we eliminate the need for a trusted workload generator. 
Because all variables, execution paths, and memory states within an Ethereum smart contract are universally public, the contract is incapable of hiding pre-computed prime factors.
To achieve this without compromising the puzzle, the generator relies on public, on-chain pseudo-randomness and deterministic execution rather than hidden inputs. 
This shifts the task of generating a transparent, mathematically sound benchmark to the decentralized network itself.
Hence, this public verifiability eliminates the need for a trusted third party, providing assurance that the generated benchmarks are devoid of hidden secrets or back doors.

\begin{figure*}
    \centering
    \includegraphics[width=0.88\linewidth]{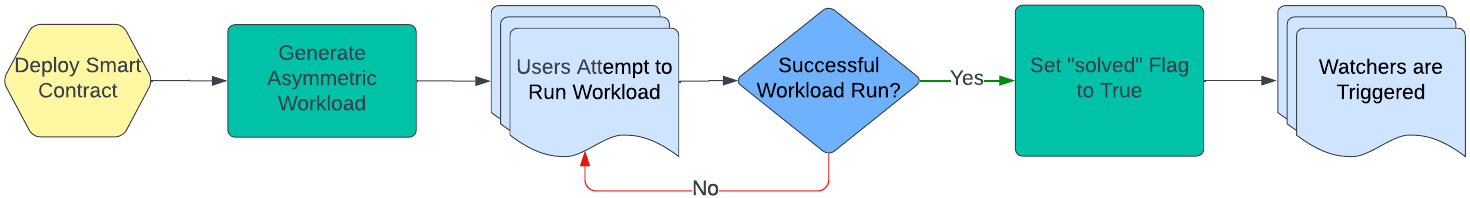}
    \caption{An outline of the automated triggers through blockchain-based workload verification. Generation is done on-chain without pre-computed secrets, ensuring no bias. ``Watchers'', here, are anything that runs after a succesful workload run.}
    \label{fig:outline}
\end{figure*}

The race to achieve quantum supremacy involves immense effort and resources, while the payout is the monumental prestige and technological leadership associated with the milestone. 
Yet, as evidenced by the cycle of classical rebuttals to previous claims, traditional self-reporting often leaves these achievements open to intense debate. 
Our framework resolves this by offering indisputable proof for such claims.
Any successful submission definitively settles the debate and permanently secures the claimant's reputation on-chain. 
Furthermore, this reputational incentive can be augmented by leveraging the native functionality of smart contracts to lock digital assets as a financial bounty. 
This creates a globally accessible honeypot that economically incentivizes an actor to submit a valid solution the moment their hardware is capable.
Together, the promise of undeniable prestige and direct financial reward ensures that the benchmark actively attracts the most advanced quantum systems available.

This decentralized benchmarking framework is not limited to measuring a single milestone but instead establishes a tunable gradient for characterizing quantum capabilities. 
By adjusting the puzzle complexity, such as the bit-length of integer factorization or creating a different puzzle entirely, the smart contract can be calibrated to detect early milestones of quantum advantage.
This gradient ultimately culminates at the definitive capability of quantum hardware to bypass classical cryptographic algorithms (such as Rivest–Shamir–Adleman (RSA)~\cite{rsa} and Elliptic Curve Digital Signature Algorithm (ECDSA)~\cite{ecdsa}), which we call \emph{cryptographic quantum supremacy} and expect to be the most computationally demanding threshold of interest.
Because the blockchain acts as an immutable public ledger, and the puzzles are generated without any possibility of solution foreknowledge, any valid solution submitted across this gradient serves as cryptographically verifiable proof of advancing quantum capability.

Beyond simply tracking progress, hosting this gradient on a decentralized ledger provides a globally synchronized timestamp for the exact moment of verification. 
While traditional benchmarking relies on publications that rigorously timestamp when a claim of quantum advantage is submitted, the subsequent validation of that claim often entails months of community debate and classical simulation attempts. 
By contrast, a smart contract enforces strict, instantaneous, and consensus-driven mathematical verification. 
When a quantum processor successfully executes a workload, the submission transaction creates an indisputable, globally agreed-upon historical record of exactly when a specific quantum capability was cryptographically proven, entirely eliminating the latency between claim and verification.

Furthermore, integrating workload characterization directly into a distributed ledger enables automated technological transitions. 
Current blockchain infrastructures rely heavily on classical signature verification~\cite{verificationschemes}, which faces an existential threat from quantum adversaries~\cite{securityRisk,securityRisk2,securityRisk3,securityRisk4,Castelvecchi2023_dz}. 
While quantum-secure algorithms (e.g., Lamport signatures~\cite{lamport}) exist, they incur significant performance penalties, drastically increasing computational overhead and transaction costs (measured in gas~\cite{gas}) on the network.
Therefore, a premature adoption of such post-quantum cryptography would drastically reduce transaction throughput and impose prohibitive financial costs on users.
Conversely, waiting for human consensus after a breach has occurred exposes digital assets to immediate vulnerability.
Our proposed smart contract utilizes its synthetic benchmark as a dynamic trigger, allowing  the network to operate on highly efficient classical cryptography until the exact moment practical quantum utility is verified.
At that point, actors watching the trigger can automatically fall back to post-quantum security protocols.
Fig.~\ref{fig:outline} outlines this process.

We have implemented and thoroughly tested \emph{BloQBench}, a cryptographic supremacy contract that probabilistically generates hard-to-factor products of primes and triggers a transition to Lamport signature verification when solved.
This proves the validity of the overall framework and provides us cost estimates for deploying such a contract, which we analyze in this work.
The artifact is available at:

\begin{center}
    \href{https://doi.org/10.5281/zenodo.14664709}{ethereum-quantum-bounty}
\end{center}

\vspace{0.05in}
\noindent 
Overall, the primary contributions of this paper are:
\begin{itemize}[leftmargin=0.25in]
    
    \item \textbf{A tunable, decentralized benchmark generator:} The design of a smart contract capable of generating classically intractable puzzles without pre-computed secrets. 
    By scaling the complexity of these puzzles, the framework provides a measurable gradient to track incremental milestones in quantum computing capabilities.
    
    \item \textbf{Trustless verification capability:} The establishment of an unbiased, universally auditable framework for proving quantum supremacy milestones, leveraging the blockchain as a public ledger for benchmarking results.
    
    \item \textbf{Automated technological transitions:} A demonstration of how this verifiable benchmarking can trigger system-level fallbacks to quantum-secure protocols, successfully managing the trade-off between execution overhead and post-quantum security.

    \item \textbf{BloQBench:} An implementation of the framework using integer factorization and Lamport signatures.
\end{itemize}

%% file: sections/background_and_motivation.tex
% \clearpage
\section{Background and Motivation}\label{sec:background}

\subsection{Quantum Supremacy Benchmarking}
A significant milestone in quantum computing is quantum supremacy, which refers to a quantum computer solving a problem that would take a classical computer an infeasible amount of time to solve~\cite{preskill2012quantumcomputingentanglementfrontier}.
The pursuit of quantum supremacy has driven the development of various synthetic benchmarks. 
Early milestones in quantum execution~\cite{firstAlgorithm,supremacyTimeline} paved the way for state-of-the-art demonstrations.
A watershed moment occurred in 2019 when Google, in partnership with NASA and Oak Ridge National Laboratory, demonstrated quantum supremacy with their Sycamore quantum processor, which performed calculations in seconds that would allegedly take even advanced supercomputers thousands of years~\cite{arute}. 
Since then, numerous experiments have utilized synthetic workloads to characterize quantum processing capabilities and assert quantum supremacy~\cite{kim,morvan2023phasetransitionrandomcircuit,Neven_2024,king2024computationalsupremacyquantumsimulation,dwaveSupremacy}.

However, verifying these benchmarking claims has proven extraordinarily challenging. 
As noted by Vazirani, ``The basic question that must be answered for any such experiment is how confident can we be that the observed behavior is truly quantum and could not have been replicated by classical means''~\cite{Yang_2018}. 
In fact, many recent claims of quantum supremacy have been successfully refuted or efficiently reproduced by highly optimized classical simulation~\cite{begušić2023fastclassicalsimulationevidence,Pednault2019LeveragingSS,Pednault_Maslov_Gunnels_Gambetta_2019b}. 
This dynamic highlights the lack of a provably trustless, universally verifiable mechanism for quantum performance evaluation.

\subsection{Integer Factorization as a Workload}
To overcome the possibility of classical replication, benchmarking methodologies must leverage tasks that are inherently intractable for classical hardware to compute but trivial to verify.
Such tasks, if solved, would have verifiable quantum supremacy~\cite{yamakawa}.
Integer factorization serves as a common example. 

RSA showcases the classical infeasibility of integer factorization, as the protocol would fail if factoring were classically possible, yet it remains the current standard method of encryption~\cite{rsa}.
However, Shor's algorithm proves that this workload can be performed quantum mechanically in polynomial time~\cite{shor_97}.
Because of this reliance on RSA's classical hardness assumption, classical spoofing is not a distinct threat to guard against, and BloQBench does not distinguish how a submitted solution was obtained.
Any efficient classical factorization would itself constitute a fundamental break of RSA.

This disparity in efficiency can aid in benchmarking the performance of quantum hardware. 
By generating puzzles with varying levels of difficulty, such as factorization puzzles of varying bit-lengths, a gradient of quantum capability can be tracked. 
This point in time marks a definitive, verifiable milestone of quantum capability to bypass current cryptographic standards such as RSA~\cite{rsa} and ECDSA~\cite{ecdsa}.

\subsection{Overhead of Post-Quantum Cryptography}
The eventual realization of cryptographic quantum supremacy poses a threat to systems relying on classical signature schemes~\cite{verificationschemes}, including modern distributed ledgers~\cite{securityRisk,securityRisk2,securityRisk3,Castelvecchi2023_dz}. 
While post-quantum cryptographic alternatives exist, implementing them introduces severe computational overheads. 

Quantum-secure classical schemes such as Lamport signatures~\cite{lamport} or lattice-based cryptography~\cite{latticeSignature} require significantly larger key sizes, increased memory footprints, and execution time compared to traditional ECDSA~\cite{gan2022,ghosh2021}.
In a blockchain environment, these high hardware and processing overheads translate into high execution costs (gas fees). 
Therefore, preemptively transitioning to post-quantum cryptography before quantum hardware is capable of breaking classical encryption results in unnecessary performance degradation and financial cost.

Note that migration to post-quantum cryptography is ultimately inevitable regardless of any particular benchmarking mechanism.
The open question is \textit{when} systems must transition.
The trade-off between security and cost forms the primary motivation for an automated transition mechanism.
By utilizing the blockchain to host an unbiased, verifiable quantum benchmark, we provide a mechanism that incentivizes the open demonstration of quantum advancement while serving as an automated trigger. 
Incentives will be based on both public clout for the solver as well as a monetary bounty provided as a reward.
This allows systems to continue utilizing high-performance classical cryptography until practical quantum utility is verified, delaying the transition to costly post-quantum cryptography until necessary.
This delay offsets the cost of the trigger with the savings from avoiding premature migration.

%% file: sections/2result.tex
\section{Implementation using Factorization and Lamport Signature Verification}\label{sec:results}

This paper selects prime factorization as the best-suited puzzle to exemplify the stated goals (see Sec.~\ref{subsec:puzzle-alternatives}).
Prime factorization, the task of decomposing a positive integer $n$ into the set of prime numbers whose product is equal to $n$, is a computational challenge widely believed to be intractable for classical computers as the size of the factors increases.
However, Shor’s algorithm demonstrates an efficient quantum solution~\cite{shor_97}, and since both RSA and ECDSA rely on the difficulty of this problem, it serves as an ideal test for cryptographic quantum supremacy.
Hence, if this challenge is generated on a blockchain, where the solution is never known throughout the process of generation, then if someone were to provide a solution that is also verified on the blockchain, this would be virtually indisputable proof that cryptographic quantum supremacy has been achieved.

The contract would use techniques based on Sander's method of probabilistically generating a hard-to-factor integer~\cite{sander_eff_acc}, elaborated on further in Sec.~\ref{sec:discussion}.
This paper refers to these generated integers as locks because providing solutions (prime factorizations) to these will unlock bounty funds held in the contract.
We conclude that the contract should generate 119 locks of \modulusBitSizeWithCommafsze\ bits in order to have confidence that at least one of them will be infeasible to solve by classical means alone.
The confidence level for which we aim is about a one-in-a-billion chance that this fails.
Though we have confidence that at least one will be infeasible, the expectation, discussed in Sec.~\ref{sec:discussion}, is that 19 of them will be infeasible.

The bounty funds serve as an incentive for solvers.
Participants may attempt to provide a prime factorization for any single lock at any time, where quantum-driven solutions are likely relayed via a classical client.
If the solution is correct, then the lock is marked as solved, and no additional attempts to solve that particular lock will be allowed.
This process concludes when the final lock is solved,
at which point all funds held within the contract will be transferred to the final solver, a publicly readable flag will be set to indicate that the contract has been successfully solved, and subsequent interactions, such as new submissions or fund deposits, are disabled.
There will be an enforced minimum bounty, ensuring that computational costs of claiming the bounty are covered, requiring funds sufficient for at least \bountyGasfsze\ gas.
For a rough-estimate example, if the current market price is 23.8 Gwei per unit of gas, one Gwei being one billionth of one ETH, the contract should have at least 19.04 ETH as a bounty.
This figure represents the total cost of solving all 119 locks across the full lifetime of the contract.
However, a single lock verification costs only \costSolvePrimeFactorizationOnefsze\ gas.
As discussed later in this section, classically easy locks are expected to be solved incrementally by the community, distributing the aggregate cost rather than concentrating it on any single actor.

The contract accommodates fluctuations in gas prices by accepting unrestricted donations from any user. 
This allows interested parties to monitor gas price conditions externally and contribute additional funds as needed, helping ensure that the payout remains greater than the cost to solve the puzzles. 
Automated, on-chain bounty adjustment is a natural extension of this design but is outside the scope of this work. 
Furthermore, the reward pool is not limited to covering gas costs alone, and donations of any amount can make the solving process more lucrative.

\begin{figure*}
    \centering
    \includegraphics[width=0.9\linewidth]{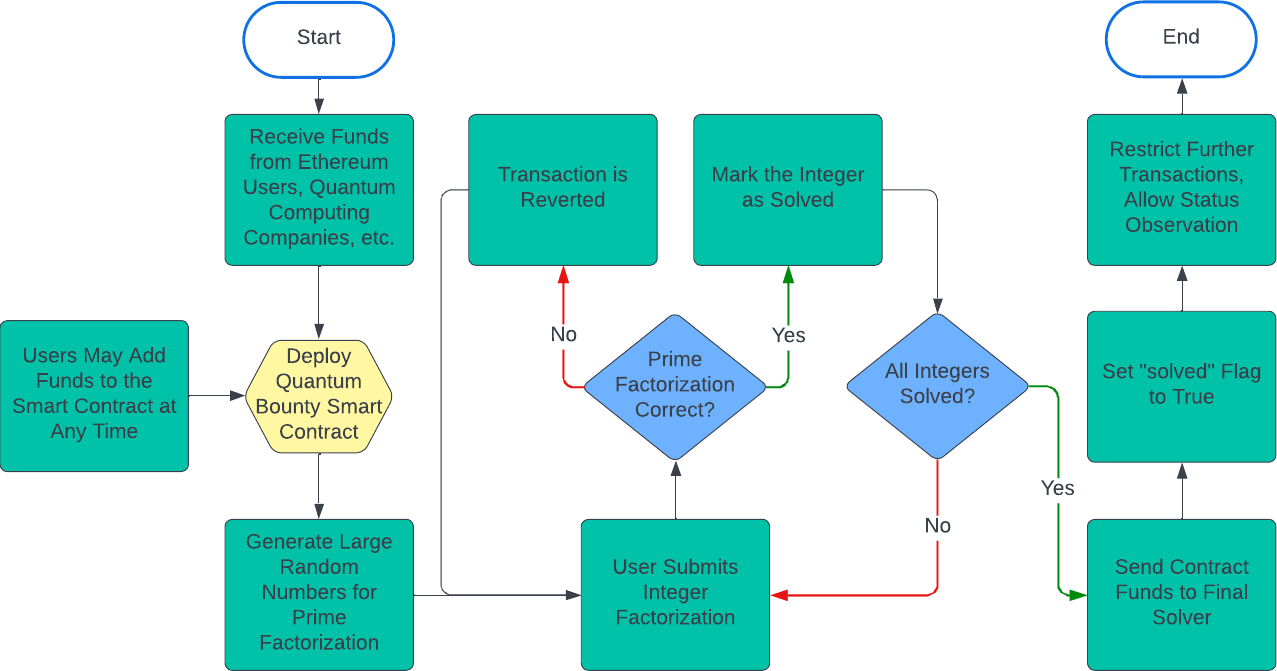}
    \caption{Flowchart of quantum bounty smart contract.}
    \label{fig:flowchart}
\end{figure*}

Fig.~\ref{fig:flowchart} illustrates the overall process of the contract.
BloQBench inherits blockchain-level operational conditions such as network congestion or gas price fluctuations from the underlying blockchain.
These conditions are not depicted as they do not affect the correctness, trustlessness, or security of the process shown.

Deployment of the contract follows the Singleton pattern, ensuring that a single instance across all network instances generates and manages the locks~\cite{erc_2470}.
After deployment, a public method that partially generates the locks must be repeatedly invoked until all locks are generated.
This is to limit the cost of each transaction so that generation happens in small pieces, both preventing a reversion of the transaction due to exceeding the gas limit, which is currently 30 million gas for Ethereum~\cite{gasLimit}, and allowing the distribution of cost over many participants rather than burdening it all onto a single user.
The completion of the generation will be marked with a flag, at which point no further invocation of the generation method will be processed, and the function accepting solution submissions will be enabled.

Known blockchain attacks must be accounted for~\cite{reentrancy,reentrancy2,solidityAttackReview,solidityAttackMetana,commitReveal}. 
Among these is front-running, a malicious technique which, with respect to this contract, would mean that a user would watch for an incoming solution, steal the solution, and submit it themselves via a transaction with a higher gas price, which users may dictate upon submission~\cite{commitReveal}.
As transactions are not necessarily committed chronologically, but rather transactions with higher priority fees are usually prioritized, users can jump the line by paying a higher fee~\cite{gas}.
To prevent front-running and maintain fairness, the submission of solutions follows a commit-reveal scheme with a one-day buffer~\cite{commitReveal}.
That is, solvers must first submit an encrypted message derived from the proposed factorization and their sending address (an address, in the context of a blockchain, is essentially a unique name representing a user and their funds).
This message's hash value is stored in the contract.
After a mandatory waiting period of one day, solvers can reveal their solution by submitting the original data used to compute the hash.
The contract verifies the solution by checking that the submitted data hashes to the stored value.

Since the sending address is included in the encryption, an attacker cannot simply submit their own copy of the original transaction of the encrypted data.
Instead, they must wait for the reveal made by the original solver, then generate a new encryption using the revealed prime factorization and their own address, and only then attempt to submit their own transaction.
To succeed, the attacker would need to flood the transaction pool, called the mempool~\cite{mempool}, with highly priced transactions for an entire day, preventing the original reveal from being committed, and then submit their own reveal before the original can be processed.
This delay is infeasible for attackers to overcome, as discussed in Sec.~\ref{sec:discussion} and visualized in Fig.~\ref{fig:commit-reveal}.

Consequently, external accounts can reliably monitor BloQBench to trigger post-quantum security workflows. 
As a demonstration, we implemented an account that dynamically adjusts its signature verification scheme for inbound transactions.
Upon receiving a transaction, the account queries BloQBench to inspect its state.
Before the benchmark is solved, signatures are verified using standard classical schemes, whereas a post-quantum Lamport signature scheme is enforced once the puzzle is solved~\cite{lamport}.

By combining blockchain security with economic incentives, this contract offers a trustless and transparent mechanism for demonstrating cryptographic quantum supremacy while providing monetary incentives to do so, which would be in addition to the societal incentives that organizations already have to prove cryptographic quantum supremacy.
Furthermore, this contract allows instantaneous fallback safety for Ethereum users to protect their funds once this cryptographic quantum supremacy has been detected.
Table~\ref{tab:summary} summarizes the parameters of this contract.
\begin{table}[ht]
\caption{Summary of quantum bounty contract parameters.}
\begin{center}
\begin{tabular}{|c|c|}
\hline
\textbf{Parameter}&\textbf{Value} \\
\hline
\text{Bounty puzzle} & \begin{tabular}{c}
     Prime Factorization \\
     via RSA-UFO generation
\end{tabular} \\
\hline
\text{Bit size of primes} & \bitSizeOfPrimesfsze \\
\hline
\text{Number of locks} & 119 \\
\hline
Bit size of locks & \modulusBitSizeWithCommafsze \\
\hline
Chance of being insecure & $10^{-9}$ \\
\hline
\text{Estimated gas to solve} & \bountyGasfsze\ \\
\hline
\text{Commit-reveal time gap} & 1 day \\
\hline
\end{tabular}
\label{tab1}
\end{center}
\label{tab:summary}
\end{table}

%% file: sections/3discussion.tex
\section{Security Evaluation}\label{sec:discussion}

\subsection{Choosing the parameters}

The bit-length of the generated integers was chosen to be the easiest to solve by quantum means while being infeasible for classical computers.
The work of Sander proves that it is possible to generate difficult-to-factor numbers, known as RSA-UFOs, without prior knowledge of their factorization~\cite{sander_eff_acc}.
This is essential in order to allow trustless generation, as any generation requiring secret knowledge on a blockchain can be potentially exploited.
An analysis by Anoncoin, another cryptocurrency using Sander's techniques, determined that the probability of generating an RSA-UFO as approximately 0.16~\cite{Anoncoin}.
Therefore, the calculation \(\log(10^{-9}) / \log(1 - 0.16) \approx 119\) determines that this contract should generate 119 locks, providing a high confidence level that at least one secure lock will be generated.
Specifically, there is about a one-in-a-billion chance that all locks can easily be factored.

Note that Theorem 1 of the Sander paper states that, letting $\xi \in (\frac{1}{3}, \frac{5}{12})$, the number of integers $\leq x$ that have two distinct prime factors $\geq x^\xi$ is $x(\frac{1}{2} \ln^2(\frac{1}{2\xi}) + O(\frac{1}{ln(x)}))$~\cite{sander_eff_acc}.
We can see that, since $O(\frac{1}{ln(x)})$ approaches $0$ as $x$ approaches $\infty$, the probability $\frac{1}{2} \ln^2(\frac{1}{2\xi}) \approx 0.082$ when $\xi = \frac{1}{3}$.
As this differs from Anoncoin's results by only a factor of approximately 2, we choose the more cost-friendly result of generating 119 locks while being confident that at least one secure lock will be generated.

We now predict the time-scale and resource requirements of solving this puzzle both classically and quantumly.
Classical security for 3072-bit RSA keys is expected to remain robust beyond 2030, while lower-sized keys may be broken before then~\cite{kaliski2003twirl}.
We choose this sized key because the timeline for quantum supremacy is uncertain and may very well not be achieved before 2030.
The most prominent reason to expect this is that existing and near-term quantum computers are noisy and error-prone, classifying them as Noisy Intermediate-Scale Quantum (NISQ) computers~\cite{nisq}, which have limited reliability, utility, and scalability. 
While NISQ systems show promise in outperforming classical computers for certain applications, such as optimization~\cite{farhi2014quantum,farhi2016quantum}, 
their capability in integer factorization remains restricted to numbers of only a few tens or, at most, a few hundred bits~\cite{ayanzadeh2020reinforcement}—far from practical use cases.
Quantum error correction holds the potential to enable fault-tolerant quantum computing~\cite{errorCorrection}, allowing reliable execution of quantum programs on noisy devices. 
However, achieving large-scale fault-tolerant quantum computing necessitates significant technological breakthroughs. 
In quantum error correction, for instance, a logical qubit is encoded across multiple physical qubits, often requiring hundreds or thousands of them, making near-term implementation impractical. 
Even with sufficient physical qubits, performing logical operations and executing real-time error detection and correction, such as in superconducting qubits within one microsecond timescales, remains a major challenge~\cite{alavisamani2024promatch}.  
Hence, while breaking 256-bit elliptic curve encryption would require approximately 2,330 logical qubits~\cite{Roetteler}, achieving this feat within one day would require an estimated \(13 \times 10^6\) physical qubits under the current fault-tolerant regime~\cite{Webber2022_yb}.

However, RSA-UFO generation requires generating random numbers three times the size of the desired large prime factors.
Desiring a key of at least 3,072 bits comprised of at least two primes, we want primes of at least $3072 / 2 = \bitSizeOfPrimesfsze$ bits each.
Hence, the contract should generate 119 locks of $\bitSizeOfPrimesfsze * 3 = \modulusBitSizefsze$ bits for a confidence of $10^{-9}$ that at least one lock cannot be factored by classical means alone.

The bounty mechanism aims to ensure that solvers are reimbursed for the computational costs associated with verifying puzzle solutions.
The expected number of prime factors of a number $N$ is on the order of $\log(\log(N))$, so the expected number of prime factors of a \modulusBitSizefsze-bit integer is $\ln(\ln(2^{\modulusBitSizefsze})) \approx 8$~\cite{numberOfPrimeFactors}.
To estimate these costs, we deployed the contract with 119 locks, each being a \modulusBitSizefsze-bit integer of known factorization.
However, each also has 16 factors instead of the expected 8 to obtain a more conservative estimate.
The cost for solving a single lock was measured at \costSolvePrimeFactorizationOnefsze\ gas, with the vast majority of this expense arising from the Miller-Rabin primality test, used to verify that the provided factors are indeed prime~\cite{millerRabin}.
Providing correct solutions to all of these locks yielded a total gas cost of \costSolvePrimeFactorizationAllfsze\ (see Sec.~\ref{sec:costs}).
Thus, since gas cost can vary depending on various factors such as the particular state of the blockchain and the equipment running calculations to put transactions on the blockchain, a bounty of at least \bountyGasfsze\ gas should be funded to the contract to ensure enough incentive.
This amount serves as the minimum viable incentive, with adjustments made dynamically to reflect fluctuations in the Ethereum gas market.
To further enhance incentive, it is recommended that the bounty be set at least many multiples over the estimated cost at current gas prices fluctuating with market conditions.

Given the high cost of providing solutions to all locks, which would create a high barrier for any solver, we decided to implement the contract so that any one lock can be solved at any time, reducing the cost per transaction.
Since we are confident that at least one lock will be infeasible to factor classically, the community can collectively solve easy locks so that only difficult locks remain.
Hence, this ensures that solving the final lock is a satisfactory completion and proof of cryptographic quantum supremacy.
Furthermore, if one did have a powerful enough quantum computer to solve the locks, they could solve all remaining difficult locks at once, preventing the issue of not being rewarded for solving a difficult lock that is not the last to be solved.

An additional possibility is to have a gradient payout scheme.
This would entail rewarding the solver every time a lock is solved.
Since we expect that easily solvable locks will be solved first, early solutions can be rewarded with a small amount of the bounty.
This would at least serve to allow the solver to recoup some of their costs to solve puzzles that do not solve the contract.
However, since we cannot know how many locks will be classically feasible or infeasible, it is uncertain how the gradient (possibly even being flat) payout could be structured in order to keep enough of a reward for the final solver.
Hence, we consider the solving of easy factorizations to essentially be part of the initial funding to deploy, sourced by the community.

\begin{figure*}[h]
    \centering
    \includegraphics[width=0.9\linewidth,height=29em]{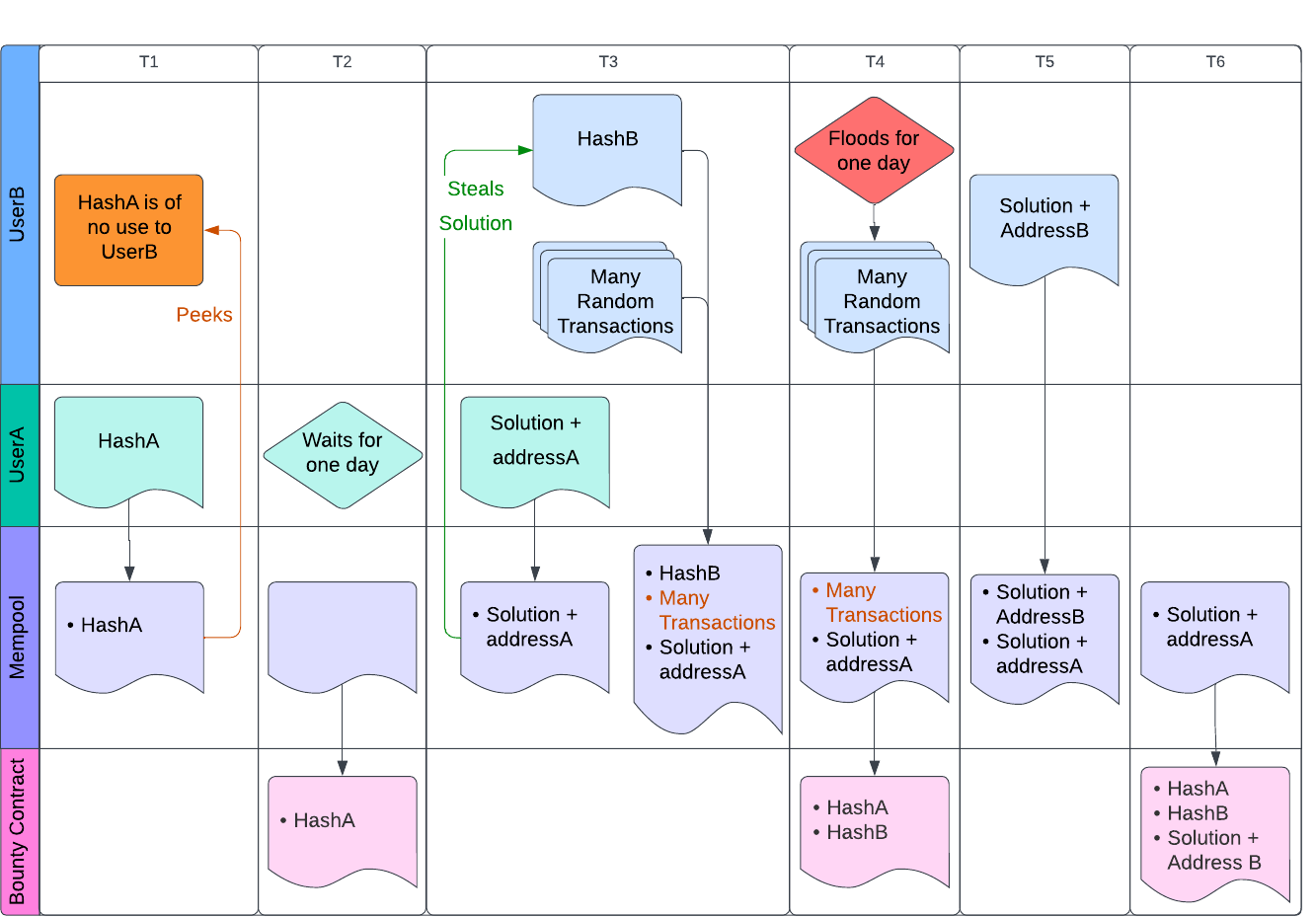}
    \caption{Timeline of the necessary steps to carry out an attack through the commit-reveal scheme.
    The infeasible step is shown in red, showing how the scheme prevents front-running. 
    Text is data currently owned by the party labeled on the left.
    Arrows into the memepool are submitted transactions. 
    Arrows into the bounty contract are committed transactions.
    HashA is generated by encrypting the solution along with the address of userA, and similarly for HashB.}
    \label{fig:commit-reveal}
\end{figure*}

The timescale for the commit-reveal scheme was chosen to be a one-day delay between commit and reveal.
This is due to the fact that an attacker would have to flood the chain to prevent the reveal long enough for said attacker to submit their own, stolen solution.
However, assuming the reveal transaction is willing to pay market rate for transaction fees, the EIP-1559 fee mechanism~\cite{eip1559} and its exponential adjustment makes it infeasible for an economic attacker to spam costly transactions to artificially increase the base-fee for an extended period of time.
Additionally, even if a large percentage of the proposers collude to censor, the inclusion of the reveal transaction on chain will be delayed but only as a function of the ratio of censoring to non-censoring proposers.
For example, if 90\% of proposers censor, then the reveal transaction will take approximately 10 times as long as expected to be included, resulting in a delay on the order of 120 seconds given mainnet block times.
If, instead, 99\% of proposers censor, then the transaction will take approximately 100 times as long, resulting in a delay on the order of 1200 seconds.
Still in these extreme regimes, reveal times on the order of a day are safe. Fig.\ref{fig:commit-reveal} illustrates a possible commit-reveal attack and the breaking point for how a one-day delay prevents this.

\subsection{Choosing the puzzle}\label{subsec:puzzle-alternatives}

Several alternative puzzles were considered before selecting the integer factorization problem, each evaluated based on its computational cost, security, and suitability for a trustless blockchain environment. 
Beyond these puzzle-level alternatives, no fully trustless and verifiable benchmarking framework is known to the authors, leaving no external prior system against which BloQBench can be compared.

One option considered was order-finding, which involves identifying the smallest positive integer \(k\) such that \(a^k \equiv 1 \pmod{n}\), given positive integers \(n\) and \(a\) coprime to \(n\), where \(n\) is referred to as the modulus and \(a\) is referred to as the base.
Since order-finding can be reduced to factoring and vice versa~\cite{WOLL1987167}, this approach would require the generation of hard-to-factor numbers as moduli and random bases coprime to these moduli.
We compare the costs of the order-finding problems with prime factorization in two ways: verifying solutions and deploying the contract.
Verifying solutions for an order-finding puzzle (see Sec.~\ref{sec:costs}) resulted in gas costs ranging from \costSolveOrderFindingMinfsze\ to \costSolveOrderFindingMaxfsze\ per iteration.
While this can be comparable to prime factorization, it is generally more costly due to the fact that the maximum order can reach a high limit due to the large key size.
We can use an estimate of \costDeployOrderFindingfsze\ gas to deploying an order-finding contract, while deploying a prime factorization contract has only a \costDeployPrimeFactorizationfsze\ gas estimate (See Sec.~\ref{sec:costs}).
Hence, we have selected prime factorization over order-finding due to order-finding's higher deployment cost and comparable solution verification cost in the average case.

Another option was requiring solvers to sign a message without access to the private key.
However, since quantum computers cannot reverse standard hash functions, one could not sign a message with the public address alone.
Hence, random generation of a public address is insufficient.
Furthermore, generating a private key along with the public address, as is usually done, introduces trust issues, as the contract deployer could access the private key and solve the puzzle without a quantum computer, undermining the trustless design.

Instead of generating an RSA-UFO, we considered alternatives such as generating large primes and their product (as in standard RSA key generation) or using a proof of quantumness based on cryptographic methods and randomness~\cite{brakerski}.
However, such protocols rely on a trapdoor, or information that must remain secret from the verifier.
Once again, the requirement of secret information during generation is incompatible with a trustless blockchain environment.

Sampling problems, as surveyed by Harrow and Montanaro, were another possibility~\cite{harrow}.
These problems aim to demonstrate quantum supremacy through the generation of samples from specific probability distributions.
However, these problems are considered unverifiable, meaning it is not possible to verify that the samples are indeed sampled from the intended probability distribution.
As the selected puzzle for this contract must be verifiable to prove correctness, these sampling problems were deemed unsuitable.

Decentralized trusted setups, where a group collectively generates data to be used as a form of proof of verification, were also considered~\cite{decTrust}.
In these setups, at least one honest participant discarding their data is enough to prevent fake proofs.
While this indeed uses minimal trust assumptions, this approach does not achieve full trustlessness.
Further exploration may reveal decentralized setups with favorable trade-offs in cost or performance, but this is left for future work.

Finally, ``Verifiable Quantum Advantage without Structure'', as proposed by Yamakawa and Zhandry, appears promising~\cite{yamakawa}.
Their method uses parameters for Folded Reed-Solomon codes~\cite{frs,improvedFrs}, offering a simpler setup with potentially lower deployment and solution verification costs.
Moreover, this puzzle might require fewer qubits to solve compared to prime factorization, allowing it to serve as an early warning system rather than a confirmation of already vulnerable security.
However, the exact parameters needed for detecting blockchain security vulnerability are unclear at the moment and left for future work.

%% file: sections/4methods.tex
\section{Evaluation Methodology}

\subsection{Quantum bounty contract implementation}

All implementations were carried out using Solidity and accompanied by extensive testing, using Hardhat.
Firstly, a fundamental building block for the proposed puzzles is a random bytes accumulator~\cite{randomBytesAccumulator}.
This accumulator contains a main function that accepts a 256-bit integer and appends it to an accumulation of bytes, and the function is meant to be called repeatedly until the required number of bytes for a lock is reached.
This process continues until the desired number of locks has been generated.
Tests included verifying that the accumulator continues to accept additional random bytes while insufficient bytes have been generated, halts accumulation once the target is achieved, and discards excess bits that exceed the target~\cite{randomBytesTests}.

Next, a general bounty contract was implemented to manage the bounty and solution verification process~\cite{bountyContract}.
This contract provides core functionality, such as maintaining a publicly accessible flag to indicate whether the bounty has been solved, offer publicly readable locks after generation, allowing contributions to additional bounty funds, and transferring funds upon solving the puzzle.
Generation of locks and verification of solutions are left to concrete subcontracts that implement the specific generation and verification processes.
Tests validated essential behaviors, including awarding the bounty to the solver and setting the solved flag upon correct solution submission, prohibiting further bounty funding or solution attempts after the bounty has been solved, and ensuring no state changes occur when incorrect solutions are provided~\cite{bountyContractTests}.
Commit-reveal tests confirmed users can retrieve or replace their commits, early reveal attempts before the required one-day waiting period are prevented, and new commits after the contract has been solved are disallowed.

The prime-factoring bounty contract was designed to function independently of the method used to generate products of primes~\cite{primeFatoringBounty}.
This contract, being a subcontract to the general bounty contract, implements the solution verification process while serving as a superclass for lock generation implementations, including the RSA-UFO generation contract.
Testing confirmed that solutions were only accepted when all provided factors for a given puzzle were prime~\cite{primeFactoringTests,millerRabinSolTests}, using the Miller-Rabin primality test~\cite{millerRabin,millerRabinSol}, and their product matched the lock value.

The main RSA-UFO generation contract represents the final building block of the implementation~\cite{primeFatoringBountyRsaUfo}.
This contract is deployed with parameters specifying the number of locks and the number of bytes per lock.
Its lock generation implementation is a function that invokes the random bytes accumulator. 
Called repeatedly, each time it supplies a new, random 256-bit integer as input.
To comply with Ethereum's transaction cost limits, this function requires multiple invocations rather than a single call to accrue bytes, ensuring that lock generation transactions remain within permissible gas limits~\cite{gasLimit}.
The deployment and verification costs reported in Section~\ref{sec:costs} directly reflect this partitioned design, showing that no individual transaction involved in generating or solving the \modulusBitSizeWithCommafsze-bit locks exceeds the 30 million gas block limit.
Tests included verifying that lock generation cannot be invoked after all bytes have been accumulated, that subsequent deployments produce distinct and non-deterministic locks, and that locks are correctly generated with the specified size~\cite{primeFactoringRsaUfoTests}.

Additionally, a proof-of-concept Ethereum account that transitions to a quantum-secure verification scheme upon detection of cryptographic quantum supremacy was implemented~\cite{fallbackAccount}.
This account was tested to confirm that it uses standard signature verification until the quantum bounty flag is marked as solved~\cite{fallbackAccountTests}, after which it transitions to Lamport signatures~\cite{lamport}.
While this implementation, as written, is currently impractical due to Ethereum’s transaction cost limitations and high gas costs associated with Lamport signature verification performed in a single shot, it demonstrates the feasibility of such an approach with minor modifications for real-world applications.
While this account-level construction is not a complete blockchain migration protocol, broader system-level concerns can naturally be layered on top of the trustless trigger BloQBench provides.

\subsection{Costs of deploying and verifying: order-finding vs. prime factorization}\label{sec:costs}
Out of several implemented, tested, and evaluated contracts, the order-finding contract was evaluated as the primary alternative to prime factorization~\cite{orderFindingContract,orderFindingTests}, forming the basis for trade-off analysis in Sec.~\ref{sec:discussion}.
Deploying an order-finding contract with 119 locks with moduli of \modulusBitSizeWithCommafsze\ bits resulted in a cost of \costDeployOrderFindingfsze\ gas, which includes testing that the generated base was neither 1 nor -1 (mod $n$) and was coprime with the modulus~\cite{deployOrderFinding}.
Alternatively, deploying without checking for being coprime could also use a probabilistic method. 
Deploying the contract at 119 locks without checking for base coprimality resulted in a cost of \costSolveOrderFindingMedianfsze\ gas.
However, since two randomly generated integers have about 0.61 chance of being coprime, one would need to generate 23 random pairs to have a one in a billion chance of having no coprime pairs~\cite{collins}.
So, this probabilistic method would also cost a large amount, possibly more, depending on the satisfactory probability.
Deploying the prime factorization puzzle, on the other hand, resulted in a cost of \costDeployPrimeFactorizationfsze\ gas when generating 119 locks of size \modulusBitSizeWithCommafsze\ bits~\cite{deployPrimeFactoring}.

To garner a general idea of verifying submitted solutions, an order-finding contract was deployed with a lock having a random \modulusBitSizefsze-bit modulus and a random \modulusBitSizeMinusOnefsze-bit base~\cite{costOfOrderFinding}.
Cleve defines the quantum order-finding problem to have an order no greater than twice the bit size of the modulus, i.e. \maxOrderfsze\ bytes in this case~\cite{cleve}.
Therefore, \maxOrderfsze\ random solutions of byte size equal to its iteration were sent to the contract.
The maximum gas cost from these iterations was \costSolveOrderFindingMaxfsze\ gas, the minimum was \costSolveOrderFindingMinfsze, the mean was \costSolveOrderFindingMeanfsze, and the median was \costSolveOrderFindingMedianfsze.
By comparison, verifying prime factorization across 119 locks of \modulusBitSizeWithCommafsze\ bits~\cite{costOfPrimes} cost \costSolvePrimeFactorizationOnefsze\ gas for a single lock and \costSolvePrimeFactorizationAllfsze\ gas for all locks.

%% file: sections/5conclusion.tex
\section{Conclusion}

We have proposed and demonstrated a decentralized benchmarking framework for verifying practical quantum utility through synthetic quantum workloads.
By deploying a smart contract that generates classically intractable synthetic workloads without pre-computed secrets, we established a universally auditable, trustless methodology for proving cryptographic quantum supremacy.
Furthermore, we demonstrated, with a signature scheme proof-of-concept, how this benchmarking metric can serve as a trigger for automatically transitioning distributed ledgers to utilize post-quantum security protocols only when dangerous quantum capabilities are verified.

A key area for ongoing research within this framework is benchmark tuning, specifically the selection and calibration of the synthetic workloads.
Alternative puzzles could provide a gradient of quantum milestones, allowing systems to both track quantum progress and balance the trade-offs of automated triggers.
On one end of the trade-off spectrum, deploying lower-complexity workloads enables earlier detection of intermediate quantum advantage, though it risks triggering costly transitions prematurely.
On the other extreme, relying on highly complex workloads delays the implementation of inefficient post-quantum security, maximizing classical computing efficiency until definitive proof of cryptographic quantum supremacy is observed.

BloQBench, the integer factorization puzzle implemented in this work, represents the latter extreme.
The successful completion of this benchmark signifies that quantum capabilities have definitively surpassed classical cryptographic standards such as RSA and ECDSA.
This provides the public with a near-instantaneous trigger mechanism to optimize execution costs while ensuring post-quantum readiness when necessary.
Future work will explore alternative workloads to detect earlier, weaker forms of quantum advantage.
Regardless of the specific workload deployed, the core requirement for this framework remains that the benchmark must be classically verifiable and generated without hidden parameters.
This framework then establishes a transparent, unbiased, and assured metric for quantum performance evaluation.

%% file: sections/artifact.tex
\appendix
\section{Artifact Appendix}

%%%%%%%%%%%%%%%%%%%%%%%%%%%%%%%%%%%%%%%%%%%%%%%%%%%%%%%%%%%%%%%%%%%%%
\subsection{Abstract}

BloQBench is a decentralized benchmarking framework that probabilistically generates classically intractable synthetic quantum workloads without relying on pre-computed secrets. 
Establishing a universally auditable and trustless methodology for proving cryptographic quantum supremacy requires transparent mechanisms. 
This framework implements a tunable prime factorization puzzle to trustlessly generate hard-to-factor integers.
It further introduces a capability to automatically trigger transitions to quantum-secure protocols, such as Lamport signatures, once the puzzle is solved. 
Additionally, the framework successfully incorporates a time-delayed commit-reveal scheme to prevent front-running attacks during the submission phase.
Experimental evaluations, at the time of experimentation, demonstrate that deploying a benchmark of 119 locks at 4,608 bits incurs a one-time cost of approximately 237,898,716 gas. 
Verifying a single prime factorization solution via the Miller-Rabin primality test costs approximately 6,110,290 gas, with a total gas cost of approximately 741,048,802 for providing solutions to all locks. 

\subsection{Artifact check-list (meta-information)}

{\small
\begin{itemize}
  \item {\bf Algorithm: } Prime Factorization via RSA-UFO generation~\cite{sander_eff_acc}, Miller-Rabin Primality Test~\cite{millerRabin}, Lamport Signatures~\cite{lamport}, Commit-Reveal Scheme~\cite{commitReveal}.
  \item {\bf Program: } BloQBench Solidity smart contracts.
  \item {\bf Compilation: } Solidity compiler managed via Hardhat (v2.6.6)~\cite{hardhat}.
  \item {\bf Data set: } 119 dynamically generated 4,608-bit integer locks (RSA-UFOs).
  \item {\bf Run-time environment: } Node.js (v18.17), Yarn (v1.x lockfile format), and Hardhat EVM local network utilizing \texttt{ethereumjs-util} (v7.1.5) and \texttt{ethereumjs-wallet} (v1.0.1). Tested on Linux and macOS.
  \item {\bf Hardware: } Any standard x86-64 or ARM64 workstation (minimum 8 GB RAM; no GPU or quantum hardware required).
  \item {\bf Execution: } Smart contract deployment is executed via Hardhat Deploy (\texttt{npx hardhat deploy --tags \textless TAGS\textgreater}) and targeted test suite execution is performed using Hardhat pattern filtering (\texttt{npx hardhat test --grep "\textless SUBSTRING\textgreater"}).
  \item {\bf Metrics: } EVM Gas consumption (deployment, puzzle generation, and verification gas costs)~\cite{gas}.
  \item {\bf Output: } Console gas consumption reports, EVM transaction receipts, and contract boolean state assertions (e.g., the \texttt{solved} flag).
  \item {\bf Experiments: } Hardhat test suites evaluating puzzle generation, commit-reveal front-running protections, and solution verification.
  \item {\bf How much disk space required (approximately)?: } $<$ 1 GB (including \texttt{node\_modules} installed via Yarn).
  \item {\bf How much time is needed to prepare workflow (approximately)?: } $<$ 30 minutes (\texttt{yarn install}).
  \item {\bf How much time is needed to complete experiments (approximately)?: } $<$ 30 minutes for local EVM simulation.
  \item {\bf Publicly available?: } Yes.
  \item {\bf Code licenses (if publicly available)?: } MIT License.
  \item {\bf Workflow automation framework used?: } Hardhat (v2.6.6).
  \item {\bf Archived (provide DOI)?: } Zenodo archive DOI \texttt{10.5281/zenodo.14664709}.
\end{itemize}
}

%%%%%%%%%%%%%%%%%%%%%%%%%%%%%%%%%%%%%%%%%%%%%%%%%%%%%%%%%%%%%%%%%%%%%
\subsection{Description}

\subsubsection{How to access}

The source code for the artifact is publicly available on GitHub and archived on Zenodo at \texttt{https://doi.org/10.5281/zenodo.14664709}.

\subsubsection{Hardware dependencies}

A standard computing environment (PC or laptop) with internet access is sufficient to download the repository and run the local blockchain simulation.

%%%%%%%%%%%%%%%%%%%%%%%%%%%%%%%%%%%%%%%%%%%%%%%%%%%%%%%%%%%%%%%%%%%%%
\subsubsection{Software dependencies}

The evaluation environment requires Node.js (tested with v18.17), Yarn (v1.22+), and the exact locked dependencies specified in \texttt{package.json} and \texttt{yarn.lock}.

\subsubsection{Data sets}

No external datasets need to be downloaded. The smart contract dynamically generates the 119 classically intractable 4,608-bit integers on-chain using a random bytes accumulator.

%%%%%%%%%%%%%%%%%%%%%%%%%%%%%%%%%%%%%%%%%%%%%%%%%%%%%%%%%%%%%%%%%%%%%
\subsection{Installation}

Clone the repository \texttt{10.5281/zenodo.14664709} and ensure the \texttt{quantum-bounty} branch is checked out. Ensure Node.js (v18.17+) and Yarn (v1.22+) are installed on your machine, then run \texttt{yarn install} in the project root directory to automatically set up the testing environment and install all required dependencies.

To execute deployments or run scripts on the Sepolia testnet, create a \texttt{.env} file in the project root directory and retrieve the required credentials as follows:

\begin{itemize}
  \item \textbf{Deployment Wallet Private Key (\texttt{PRIVATE\_KEY}):}
  Export the private key of a dedicated testnet wallet (e.g., exported from MetaMask or generated via CLI). Ensure the account is pre-funded with Sepolia ETH (available from public faucets such as \url{https://sepoliafaucet.com}). 
  \begin{center}
    \texttt{PRIVATE\_KEY="<your\_wallet\_private\_key>"}
  \end{center}

  \item \textbf{Infura RPC Project ID (\texttt{INFURA\_ID}):}
  Sign up for a free account at Infura (\url{https://infura.io}), create a new key within the Web3 API dashboard, and copy the API/Project ID from the project settings.
  \begin{center}
    \texttt{INFURA\_ID="<your\_infura\_project\_id>"}
  \end{center}
\end{itemize}

%%%%%%%%%%%%%%%%%%%%%%%%%%%%%%%%%%%%%%%%%%%%%%%%%%%%%%%%%%%%%%%%%%%%%
\subsection{Experiment workflow}

The paper's core claims can be validated by execution scripts detailed in the ``Testing Methodology Claims and Results'' section of the \texttt{README.md} in the repository.

%%%%%%%%%%%%%%%%%%%%%%%%%%%%%%%%%%%%%%%%%%%%%%%%%%%%%%%%%%%%%%%%%%%%%
\subsection{Evaluation and expected results}

Executing the included testing suites will yield the following expected results, which substantiate the performance and cost claims presented in the manuscript:
\begin{itemize}
    \item Deploying the prime factorization puzzle with 119 locks of 4,608 bits should incur a one-time cost of approximately 237,898,716 gas.
    \item Verifying the solution for a single lock should cost approximately 6,110,290 gas, with the majority of the cost stemming from the Miller-Rabin primality test.
    \item Providing verified solutions to all 119 locks should yield a total gas cost of approximately 741,048,802 gas.
    \item The commit-reveal testing suite should successfully demonstrate that solvers cannot bypass the mandatory one-day waiting period, effectively blocking front-running attempts.
    \item The fallback account proof-of-concept tests should demonstrate a successful transition from classical signatures to Lamport signatures immediately after the benchmark \texttt{solved} trigger is activated.
\end{itemize}

%%%%%%%%%%%%%%%%%%%%%%%%%%%%%%%%%%%%%%%%%%%%%%%%%%%%%%%%%%%%%%%%%%%%%
\subsection{Experiment customization}

The framework is designed to be tunable. Users can modify the contract deployment parameters to scale the complexity of the generated puzzles. For example, adjusting the bit-length of the integer factorization or the total number of generated locks creates a different tunable gradient for detecting earlier milestones of quantum advantage, allowing researchers to explore different security and cost trade-offs.

%%%%%%%%%%%%%%%%%%%%%%%%%%%%%%%%%%%%%%%%%%%%%%%%%%%%%%%%%%%%%%%%%%%%%
\subsection{Methodology}

Submission, reviewing and badging methodology:

\begin{itemize}
  \item \url{https://www.acm.org/publications/policies/artifact-review-and-badging-current}
  \item \url{https://cTuning.org/ae}
\end{itemize}